\documentclass[twocolumn,amsmath,amssymb,aps,prl,reprint,longbibliography]{revtex4-1}
\usepackage{graphicx}
\usepackage{xcolor}
\begin{document}

\title{
Vortex Dynamics, Pinning, and Magic Angles on Moir{\' e} Patterns
} 
\author{
Wenzhao Li$^{1}$, C. J. O. Reichhardt$^{2}$, B. Jank{\' o}$^1$, and
C. Reichhardt$^{2}$
} 
\affiliation{
$^1$ Department of Physics, University of Notre Dame, Notre Dame, Indiana 46656, USA  \\
$^2$ Theoretical Division and Center for Nonlinear Studies,
Los Alamos National Laboratory, Los Alamos, New Mexico 87545, USA\\ 
} 

\date{\today}
\begin{abstract}
We examine pinning and dynamics of Abrikosov vortices interacting with pinning centers placed in a moir{\' e} pattern for varied moir{\' e} lattice angles. We find a series of magic angles at which the critical current shows a pronounced dip corresponding to lattices in which the vortices can flow along quasi-one-dimensional channels. At these magic angles, the vortices move with a finite Hall angle. Additionally, for some lattice angles there are peaks in the critical current produced when the substrate has a quasiperiodic character that strongly reduces the vortex channeling. Our results should be general to a broad class of particle-like assemblies moving on moir{\' e} patterns.  
\end{abstract}
\maketitle

Moir{\' e} patterns are produced by the interference effects
that occur when two identical lattices are placed on top of each other and
then one of the lattices is shifted or rotated
\cite{Kobayashi96,Miao16}. 
In condensed matter systems,
such patterns
can appear in a double layer system when one layer is rotated with
respect to the other. For certain rotation angles,
large scale superlattice ordering occurs that
can strongly affect the electronic properties,
as found in bilayer graphene \cite{Shallcross10,Cao18,Yao18}.
Here we present a study of the pinning and dynamics of assemblies of particles
interacting with moir{\' e} pinning patterns. We find that, as the
moir{\' e} pattern is varied, qualitatively new transport patterns emerge
for certain {\it magic} twist angles, giving rise to enhanced longitudinal
and transverse (Hall) particle currents.

One of the most ideal systems for studying pinning and sliding dynamics
on different types of pattered substrates is 
vortices in type-II superconductors.
In this system, a variety of nanostructuring techniques can be
used to realize different
pinning array geometries including
square
\cite{Baert95,Harada96,Reichhardt98a,Metlushko99a,Grigorenko03,Berdiyorov06,Crassous11},
triangular \cite{Reichhardt98a,Berdiyorov06,Martin97},
rectangular \cite{Martin99,Karapetrov05,Reichhardt06},
diluted \cite{Reichhardt07a,Kemmler09},
quasicrystalline \cite{Misko05,Kemmler06},
frustrated \cite{Libal09,Latimer13,Ge17}, 
conformal crystal \cite{Ray13,Wang13}, and other structures \cite{Sadovskyy19}. 
The pinning and dynamics can be measured by
examining the critical current and transport curves or by direct imaging
of the vortex configurations or trajectories.
Many of the results found for vortex pinning 
and motion can also be generalized to other
particle like systems interacting with ordered substrates,
such as vortices in Bose-Einstein condensates 
\cite{Tung06},
colloidal assemblies \cite{Mangold03,Bohlein12,Vanossi12,McDermott13a},
skyrmions \cite{Reichhardt18},
and frictional systems \cite{Vanossi13}. 

In a superconducting vortex system, the 
pinning properties are typically examined as a function of
the magnetic field by varying the number of vortices
on a fixed number of pinning sites.
For vortices interacting
with a moir{\' e} pinning array, an additional parameter is important beyond
the vortex and pinning density:
the angle $\theta$ between the two lattices that make up the moir{\' e}
pattern. 
Here we examine vortex pinning and motion in a
system with moir{\' e} pinning composed from
two triangular pinning lattices
rotated by an angle $\theta$ with respect to each other.
As a function of $\theta$,
we observe a rich variety of pinning and vortex dynamics that are
associated with
dips and peaks 
in the critical current.
At commensurate angles where an ordered interference pattern appears,
the critical current exhibits a series of dips, and
the vortices flow in ordered quasi-one-dimensional channels.
At incommensurate angles, these flow channels break apart.
Along the commensurate angles,
the vortices develop a finite Hall angle
due to the guidance or locking of the vortex motion
to the moir{\' e} pattern.
We also find that for other angles, peaks
in the critical current appear
when a quasicrystalline structure forms in the pinning lattice
which strongly suppresses easy flow channeling of the vortices.

{\it Simulation and System---}
We model a system of $N_{v}$ vortices interacting with a
moir{\' e} pattern of pinning 
sites. 
The equation of motion for vortex $i$ is given by
\begin{equation} 
\eta \frac{d {\bf R}}{dt} = {\bf F}^{vv}_{i} + {\bf F}^{p}_{i} + {\bf F}^{d} + {\bf F}^{T}_{i} \ .
\end{equation}
Here $\eta=1$ is the damping coefficient 
and the time step is set to $dt = 0.008$. 
The repulsive vortex-vortex interaction force
has the
form
${\bf F}^{vv}_{i} = \sum F_{0}K_{1}(R_{ij}/\lambda){\hat {\bf R}}_{ij}$,
where $K_{1}$ is the modified Bessel function,
$R_{ij}$ is the distance between vortex $i$ and vortex $j$,
$F_{0} = \phi^{2}_{0}/2\pi\mu_{0}\lambda^3 = 8.0/\lambda^3$, and $\lambda$ is the
penetration depth which we set equal to
$\lambda=1.8$.
We consider a system of size $L \times L$ with $L=20\lambda$ and
with
periodic boundary conditions in the $x$ and $y$ directions.
The vortex density is $n_v=N_v/L^2$.
The pinning force is given by
$F^{p}_{i} = -\sum^{N_{p}}_{k=1}F_{p}R_{ik}\exp(-R^2_{ik}/r^2_{p}){\hat {\bf R}}_{ij}$ 
where we fix $r_{p} = 0.6$.     
The pinning sites are arranged in two identical triangular lattices
with a lattice constant of $1.8$, and the lattices are rotated with
respect to each other by an angle $\theta$.
We consider $\theta=0$ to $\theta=30^\circ$ in increments of $\delta \theta=0.1^\circ$.
The thermal forces arise
from Langevin kicks with the following properties:
$\langle F_{i}^{T}(t)\rangle = 0.0$ and 
$\langle F_{i}^{T}(t)F_{j}^T(t^\prime)\rangle = 2\eta k_{B}T\delta_{ij}\delta(t - t^\prime)$.
The initial vortex configurations are
obtained
by starting from a high temperature liquid state and cooling down to $0$K in 
80 intervals, where we wait $10^4$ time steps during each interval.
After annealing 
we apply a drive in the form of a Lorentz force
$F^{D} = (J \times {\bf \hat z})\phi_{0}d$ which
produces vortex motion along the $x$ direction.

We obtain the critical current by measuring the
total vortex velocity
$V_{x} = N_{t}^{-1}\sum_{t}\sum_{i}{\bf \hat x}\cdot {\bf v}_{i}$, 
where $N_{t}$ is the total number of time steps and ${\bf v}_{i}$
is the vortex velocity. When $V_{x}$ exceeds a
threshold where non-trivial steady state vortex motion occurs,
the system is defined as being depinned.  
The simulations are performed using a parallelized code, and
we typically consider 3000 configurations for each of
300 different values of $\theta$
and 10 different vortex densities.
Some representative annealed
vortex configurations for varied vortex density and $\theta$ appear
in the supplemental information \cite{Supplemental}.

\begin{figure}
\includegraphics[width=\columnwidth]{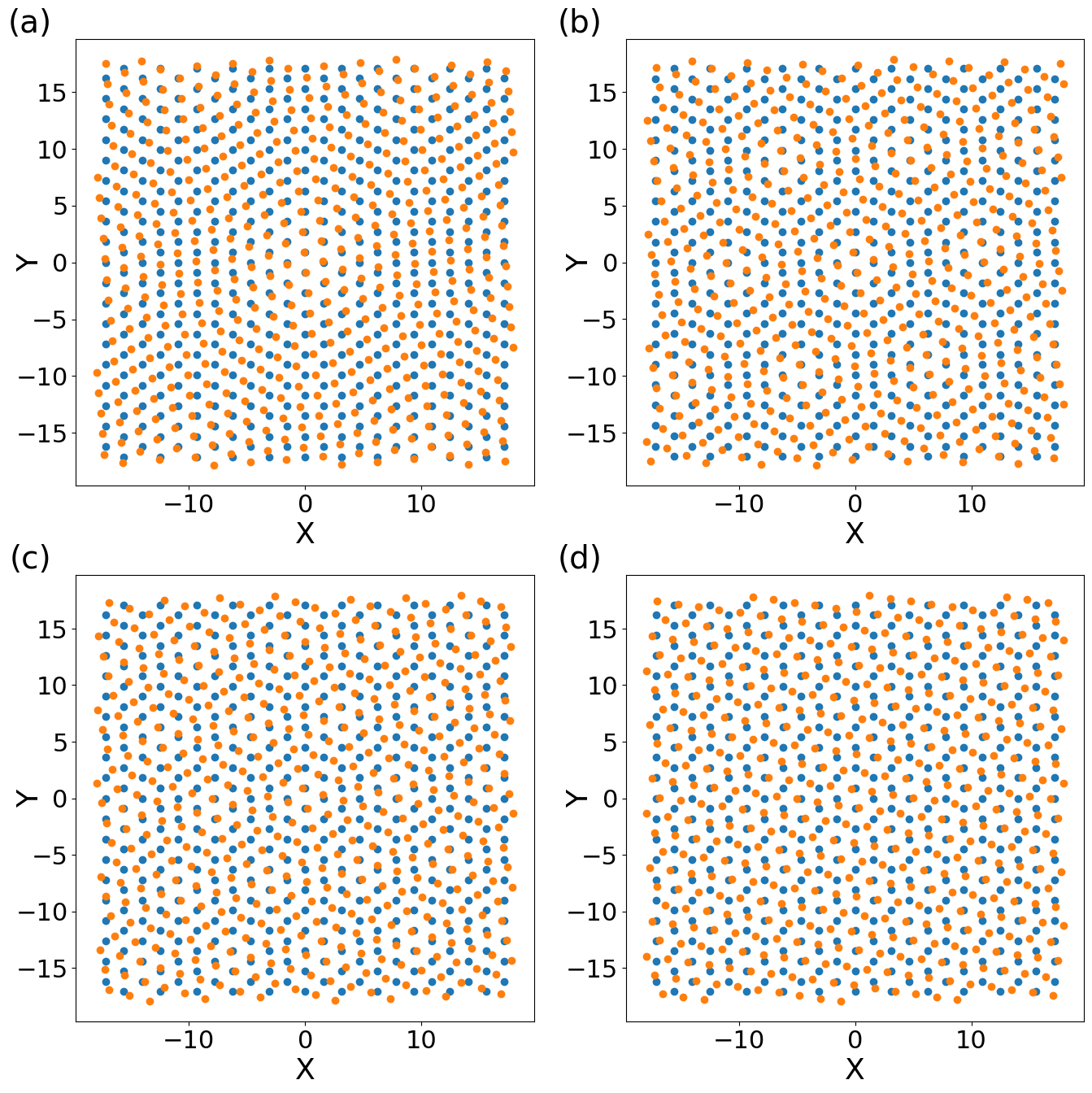}
\caption{ 
The pinning array structures for
two triangular lattices where the blue lattice is kept fixed and
the orange lattice is rotated by an angle $\theta$ of (a) 
$5.0^\circ$, (b) $9.4^\circ$, (c) $13.2^\circ$ and (d) $21.8^\circ$. 
}
\label{fig:1}
\end{figure}

{\it Results-} 
In Fig.~\ref{fig:1} we illustrate some representative
moir{\' e} pinning structures for varied angles $\theta=5.0^\circ$,
9.4$^\circ$, 13.2$^\circ$, and $21.8^\circ$
between the two lattices, which are colored blue and orange.
The pinning sites form a superlattice with a superlattice constant
that decreases as $\theta$ increases.

\begin{figure}
\includegraphics[width=\columnwidth]{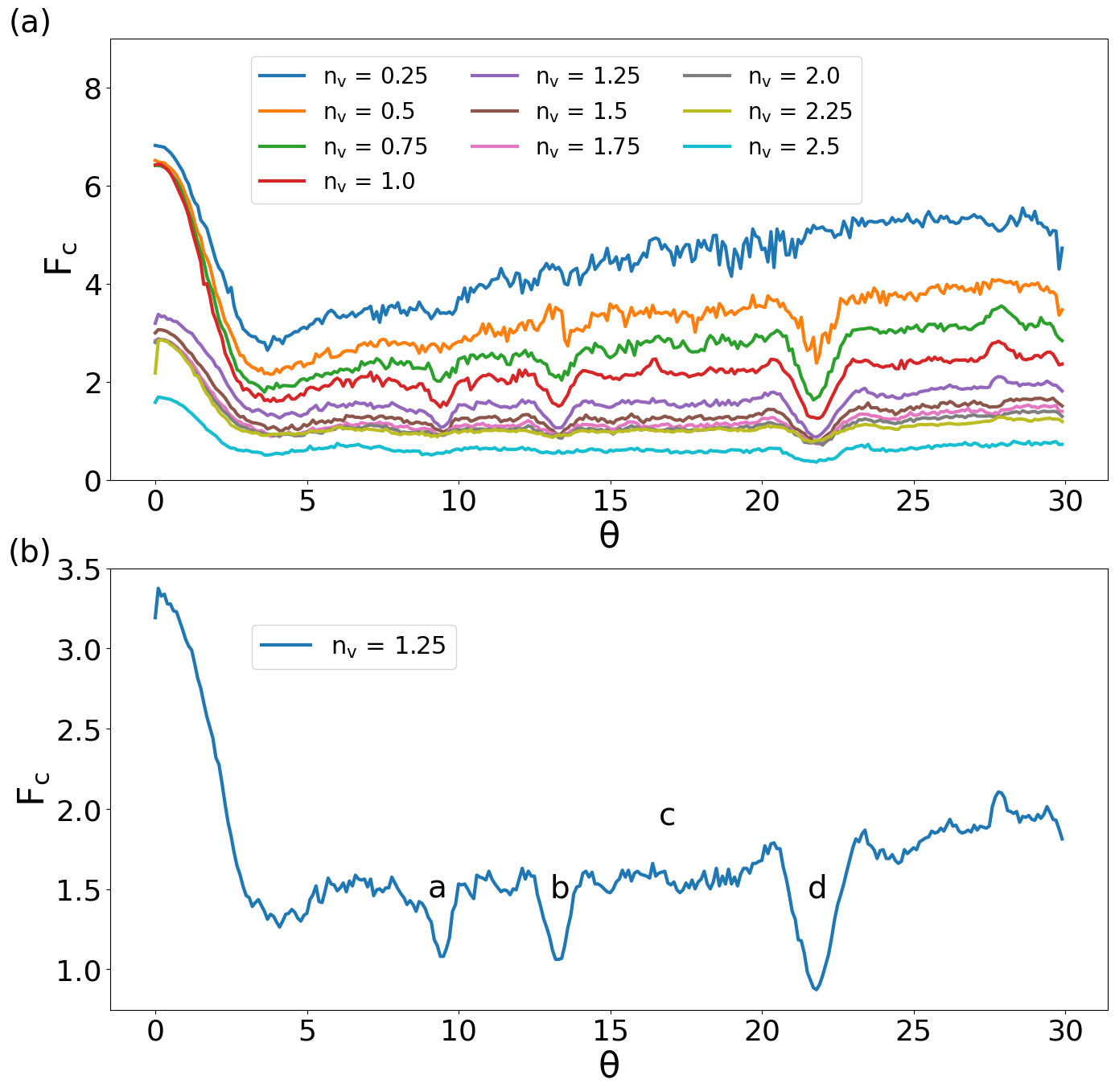}
\caption{(a) The critical current $F_c$ vs $\theta$ for the system in
Fig.~\ref{fig:1} at varied vortex densities of 
$n_{v} = 0.25$ to $2.5$ in increments of $0.25$.
(b) $F_c$ vs $\theta$ at $n_{v} = 1.25$, showing 
dips at $\theta = 9.4^\circ$, 13.2$^\circ$ and $21.8^\circ$ as well as
a peak near $28^\circ$. The letters a, b, c, and d correspond
to the locations of the images in Fig.~\ref{fig:3}.    
}
\label{fig:2}
\end{figure}

In Fig.~\ref{fig:2}(a) we plot the critical current $F_c$ versus
$\theta$ for the system in Fig.~\ref{fig:1} at vortex densities 
of $n_{v} = 0.25$ to $2.5$ in increments of $0.25$.
Here the overall critical current decreases
with increasing vortex density and
there are a series of dips at specific angles. 
The initial peak
in $F_c$ at $\theta = 0.0^\circ$ appears when
the pinning sites form a triangular lattice.  
We have also tested these results for different system sizes
and we find that the angles at which the dips and peaks occur are
insensitive to system size
\cite{Supplemental}. 

Figure~\ref{fig:2}(b) shows $F_c$ versus $\theta$
for the samples with $n_v = 1.25$, where dips
in $F_c$ appear at $\theta = 9.4^\circ$, $13.2^\circ$, and $21.8^\circ$.
In a moir{\' e} pattern formed from two triangular lattices, 
ordered or commensurate structures
occur at the following angles \cite{Yao18,LopesdosSantos12}:
\begin{equation} 
\cos(\theta) = \frac{3p^2 + 3pq + q^2/2}{3p^2 + 3pq + q^2} \ ,
\end{equation} 
where $p$ and $q$ are integers.
The values $p =1$ and $q = 1$ correspond to $\theta = 21.786^\circ$,
$p = 2$ and $q = 1$ correspond to $\theta=13.7^\circ$, and
$p = 3$, $q = 1$ corresponds to $\theta=9.4^\circ$.
The dips we observe in the critical current
match these commensurate angles.
Due to 
the symmetry of the system,
the features in Fig.~\ref{fig:2} repeat in the range
$\theta=30^\circ$ to $\theta=60^\circ$.

\begin{figure}
\includegraphics[width=\columnwidth]{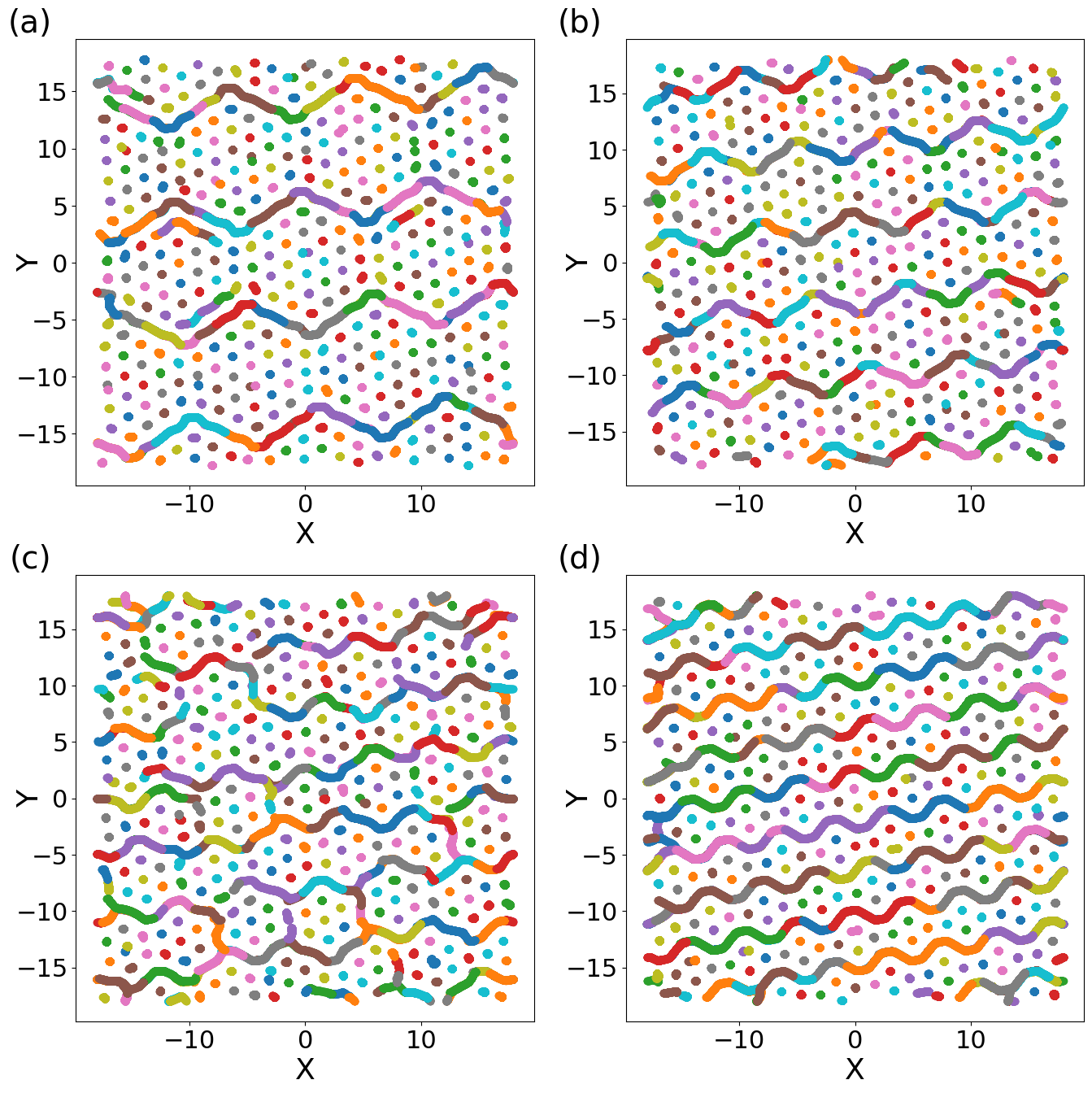}
\caption{The vortex positions (dots) and trajectories (lines)
just above depinning for the system in Fig.~\ref{fig:2}(b) with $n_v=1.25$.
Different colors indicate the motion of different individual vortices.
(a) $\theta = 9.4^\circ$ and $F_{d} = 1.5$, where quasi-one-dimensional flow
patterns form.
(b)  $\theta = 13.2^\circ$ and $F_{d} = 1.5$, with easy flow channeling.
(c) $\theta = 17^\circ$ and $F_{d} = 2.0$, an incommensurate 
angle showing more disordered channeling.
(d) $\theta = 21.8^\circ$ and $F_d=1.5$, where there is strong channeling.     
}
\label{fig:3}
\end{figure}

In Fig.~\ref{fig:2}(b), letters highlight the values of
$\theta$ at which the vortex trajectories are just able to depin,
as illustrated in Fig.~\ref{fig:3}, where the color code corresponds to
different times.
Figure~\ref{fig:3}(a) shows the trajectories
at $\theta = 9.4^\circ$ and $F_{d}= 1.5$,
where the vortices flow in
a series of quasi-one-dimensional channels
along the edges of the superlattice.
In Fig.~\ref{fig:3}(b),
at $\theta = 13.2^\circ$ and $F_d=1.5$,
a similar set of trajectories form in which
the motion follows
the superlattice
edge.
Since the superlattice
spacing decreases with increasing $\theta$,
the number of possible quasi-one-dimensional channels for motion increases
with increasing $\theta$.
In Fig.~\ref{fig:3}(c), the trajectories at a
non-commensurate angle of $\theta = 17^\circ$ and $F_{d} = 2.0$ 
are much more disordered.
At $\theta = 21.8^\circ$ and $F_d=1.5$
in Fig.~\ref{fig:3}(d),
the vortex motion again follows well-defined channels.
In general, 
the flow at incommensurate angles
has reduced channeling compared to the flow at commensurate angles.

\begin{figure}
\includegraphics[width=\columnwidth]{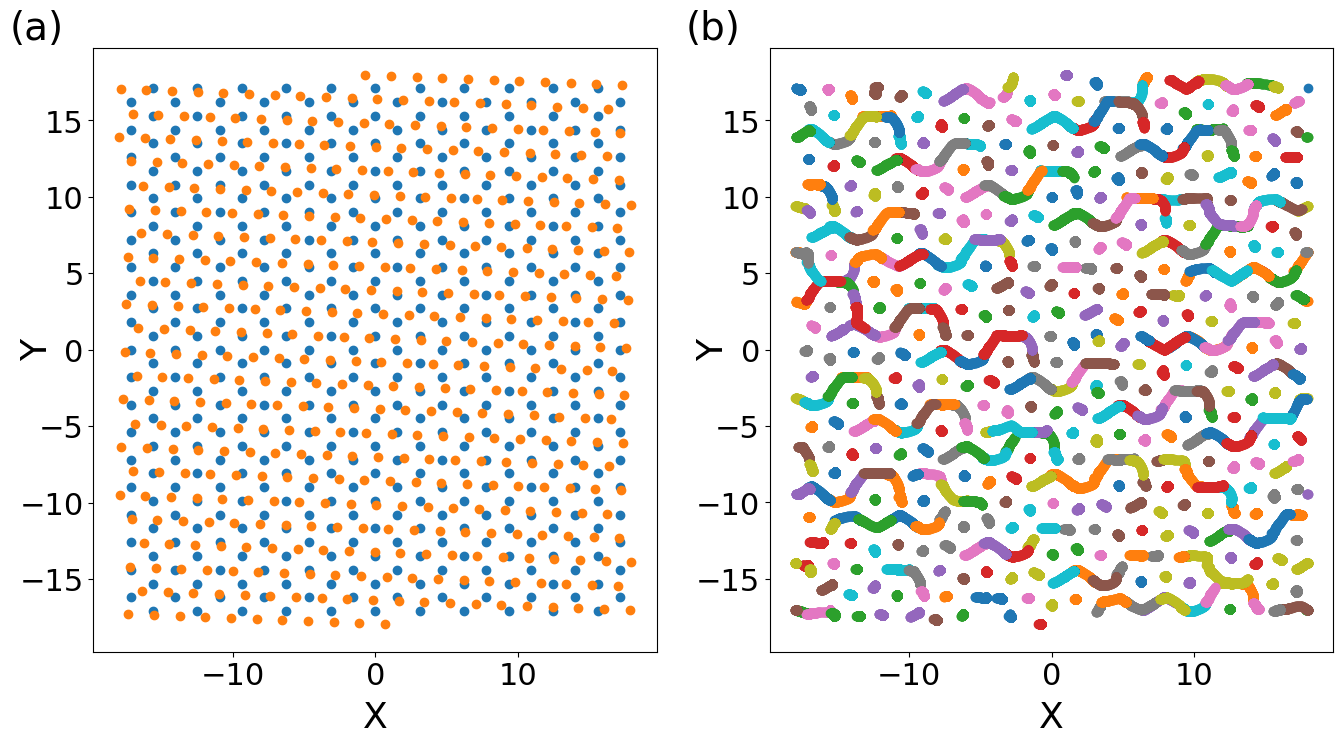}
\caption{ (a) The pinning site arrangement for the system in
Fig.~\ref{fig:2}(b) at
$\theta  = 27.9^\circ$ where a peak appears in the
critical current near $n_v=1.25$.
Here the substrate has considerable five-fold ordering or
quasiperiodic type ordering.
(b)
The vortex flow pattern over the pinning sites
at $F_{D} = 2.0$, showing a lack of ordered motion.
Different colors indicate the motion of different individual vortices.
}
\label{fig:4}
\end{figure}

Some of the peaks in $F_c$
in Fig.~\ref{fig:2}
do not correspond to commensurate angles.
The most prominent peak of this type occurs near 
$\theta = 27.9^\circ$
for vortex densities near $n_{v} = 1.25$.
In Fig.~\ref{fig:4}(a) we illustrate the pinning 
site configurations at this angle,
where we find
features such as five-fold ordering
similar to those observed in quasicrystals.
Figure~\ref{fig:4}(b) shows that the 
vortex trajectories over this substrate
just above depinning have strongly reduced channeling.
For triangular moir{\' e} patterns,
the most incommensurate angle corresponds to $\theta = 30^\circ$ \cite{Yao18}.
In our system we generally find a small dip in the
critical current when $\theta = 30^\circ$, while
the peak in $F_c$ falls at $\theta=27.9^\circ$
The downward shift of the peak location
could be a result
of the finite size of the pinning sites
or of the vortex-vortex interactions which can produce a collectively
moving state.

\begin{figure}
\includegraphics[width=\columnwidth]{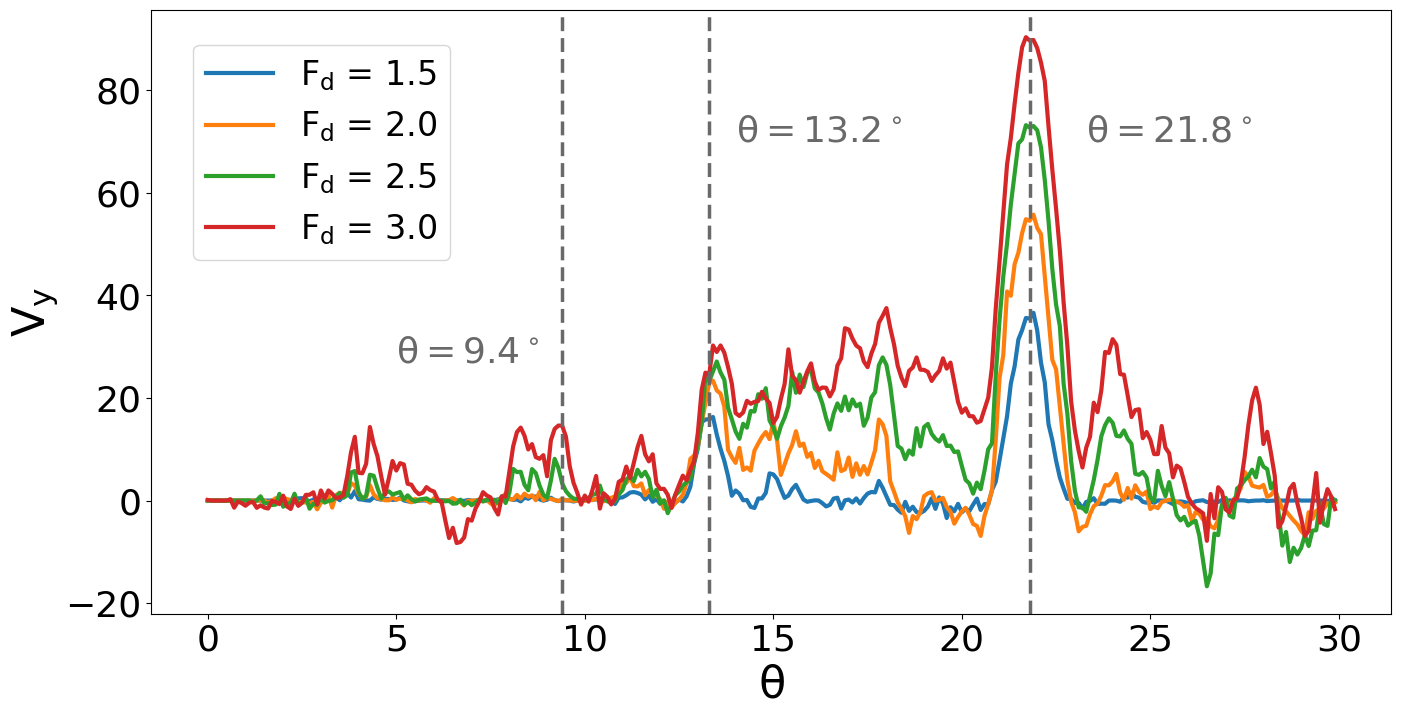}
\caption{The transverse velocity $\langle V_{y}\rangle$ vs $\theta$
for the system in Fig.~\ref{fig:2}(b) at $F_{d} = 1.5$, 2.0, 2.5, and $3.0$,
from bottom to top.
There are strong transverse velocities at the commensurate angles,
which correspond to the dips in the critical current. 
}
\label{fig:5}
\end{figure}

When vortex channeling occurs, we find
a finite Hall effect or transverse motion due to the fact that
the easy flow channels are at an angle to the driving direction,
as shown in Fig.~\ref{fig:3}.
In Fig.~\ref{fig:5} we plot $\langle V_{y}\rangle$ versus $\theta$
for the system in Fig.~\ref{fig:2}(b) 
at $F_{d} = 1.5$, 2.0, 2.5, and $3.0$.
Peaks in the Hall velocity appear at $\theta = 21.8^\circ$ and
$13.7^\circ$, with a weaker channeling effect
at $\theta = 9.4^\circ$.
There is also an extended region from
$12^{\circ} < \theta < 23^{\circ}$
in which some biased flow in the $y$ direction occurs as $F_d$ increases.
The vortex flow is generally more ordered for $\theta < 12^\circ$
even at incommensurate angles since the vortices
follow large scale zig-zag patters,
such as is shown for $\theta = 6.6^\circ$ in the SI \cite{Supplemental}.
Experimentally it is possible
to measure transverse vortex motion with various techniques
\cite{Villegas05a,Zechner18}. 
Although we find
strong variations in the critical current as a
function of the angle $\theta$,
we do not observe pronounced features as a function of field.
Instead, $F_c$ generally decreases smoothly with increasing $n_v$ except for
a jump down when the number of vortices crosses from less than to more than
the number of pinning sites \cite{Supplemental}. 

Our results could be tested using vortices on nanopatterned arrays or
for pinning sites created using multiple Bitter decorations \cite{Fasano00}.
They could also be applied to colloids
interacting with optical traps,
where it would be possible to change $\theta$ as a function of time.
Additionally, there
are proposals that the insulating state in some
bilayer systems consists of a Wigner crystal that could
undergo depinning transitions in which
the threshold could exhibit dips at the commensurate angles
\cite{Padhi18,Padhi20}. 

{\it Summary---} 
We have examined the pinning and dynamics of
vortices interacting with a moir{\' e} pattern consisting of two
triangular pinning lattices that are
rotated with respect to each other.
We find a series of dips in the critical current corresponding to
commensurate magic angles
where the system forms an ordered superlattice
and the vortices follow easy flow quasi-one-dimensional channels.
We also find that for some incommensurate
angels, the substrate has a quasicrystalline structure and
there is a peak in the critical current due to the suppression
of vortex channeling.
Dips in the critical current are correlated with
the appearance of a finite Hall angle for the vortex motion
when the channeling motion
occurs at an angle with respect to the driving direction.
Our results could be tested for vortices or colloids on moir{\' e} substrates.  

\begin{acknowledgments}
We gratefully acknowledge the support of the U.S. Department of
Energy through the LANL/LDRD program for this work.
This work was supported by the US Department of Energy through
the Los Alamos National Laboratory.  Los Alamos National Laboratory is
operated by Triad National Security, LLC, for the National Nuclear Security
Administration of the U. S. Department of Energy (Contract No. 892333218NCA000001). WL and BJ were supported in part by NSF DMR-1952841.
\end{acknowledgments}

\bibliography{mybib}

\end{document}

\title{Supplemental information for ``Vortex Dynamics, Pinning, and Magic Angles on Moir{\' e} Patterns''}
\author{
Wenzhao Li$^1$, C. J. O. Reichhardt$^2$, B. Jank{\' o}$^1$, and
C. Reichhardt $^2$
}
\affiliation{
$^1$ Department of Physics, University of Notre Dame, Notre Dame, Indiana 46656, USA\\
$^2$ Theoretical Division and Center for Nonlinear Studies,
Los Alamos National Laboratory, Los Alamos, New Mexico 87545, USA}

\maketitle

\begin{figure}
\includegraphics[width=\columnwidth]{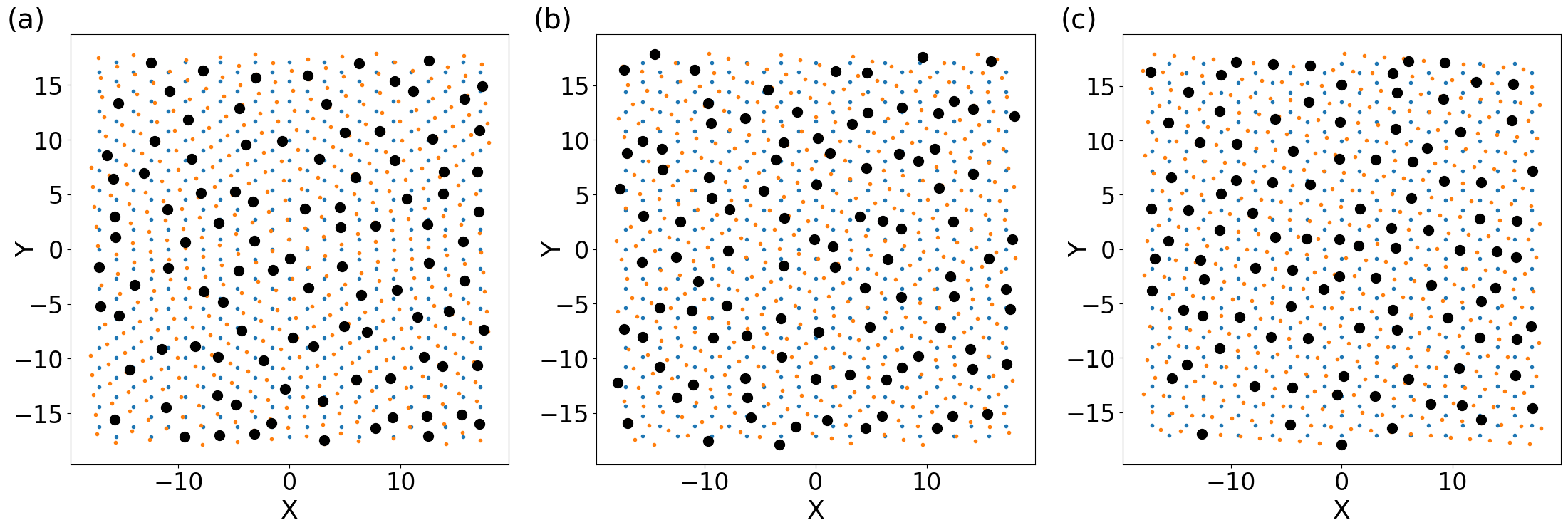}
\caption{
Vortex configurations after annealing for a system with 100 vortices (density $n_v=0.25$) for $\theta =$ (a) $5^\circ$, (b) 15$^\circ$, and (c) 25$^\circ$.
Blue dots: pinning site centers for a hexagonal lattice. Orange dots: a second hexagonal lattice rotated by $\theta$. Large dots: vortices. }
\end{figure}
\begin{figure}
\includegraphics[width=\columnwidth]{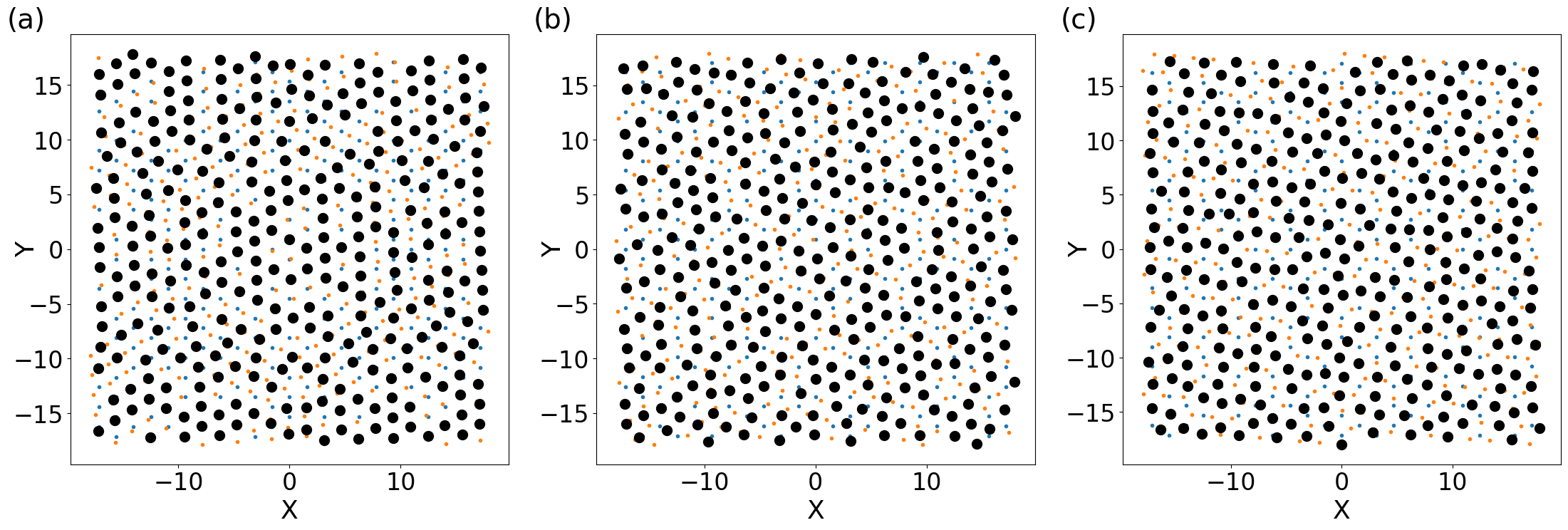}
\caption{
Vortex configurations after annealing for a system with 300 vortices ($n_v=0.75$) for $\theta =$ (a)  $5^\circ$, (b) 15$^\circ$, and (c) 25$^\circ$.
Blue dots: pinning site centers for a hexagonal lattice. Orange dots: a second hexagonal lattice rotated by $\theta$. Large dots: vortices. }
\end{figure}
\begin{figure}
\includegraphics[width=\columnwidth]{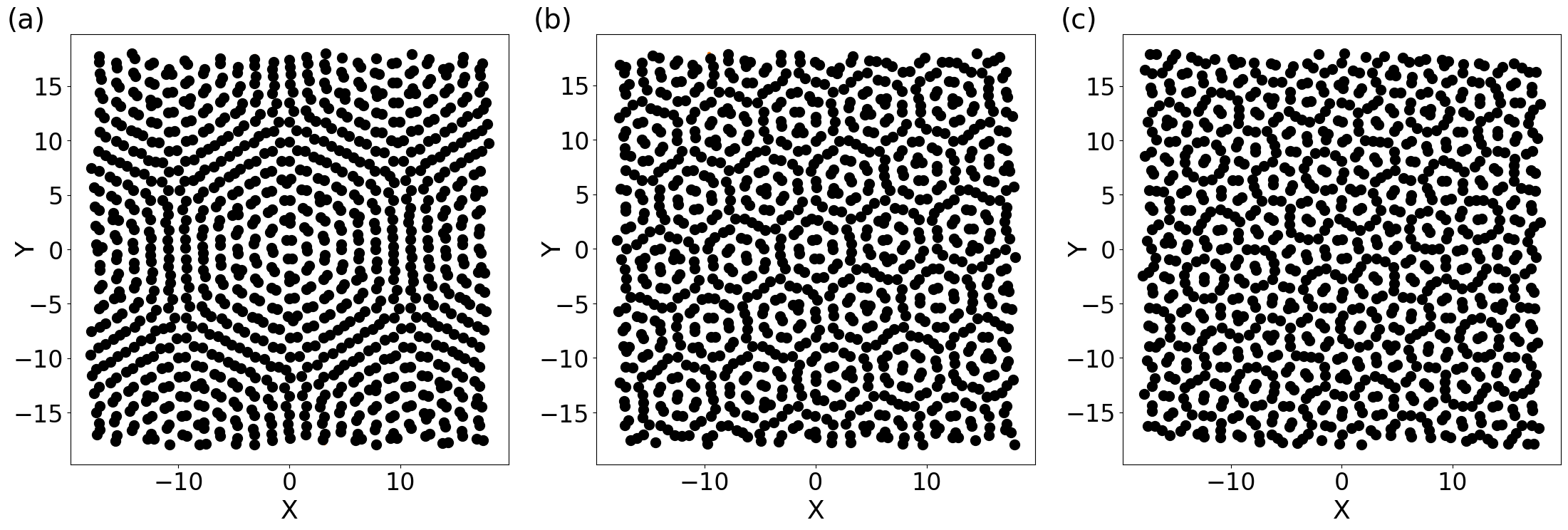}
\caption{
Vortex configurations after annealing for a system with 1000 vortices ($n_v=2.5$) for $\theta =$ (a) $5^\circ$, (b) 15$^\circ$, and (c) 25$^\circ$.
Blue dots: pinning site centers for a hexagonal lattice. Orange dots: a second hexagonal lattice rotated by $\theta$. Large dots: vortices.}
\end{figure}
\begin{figure}
\includegraphics[width=0.88\columnwidth]{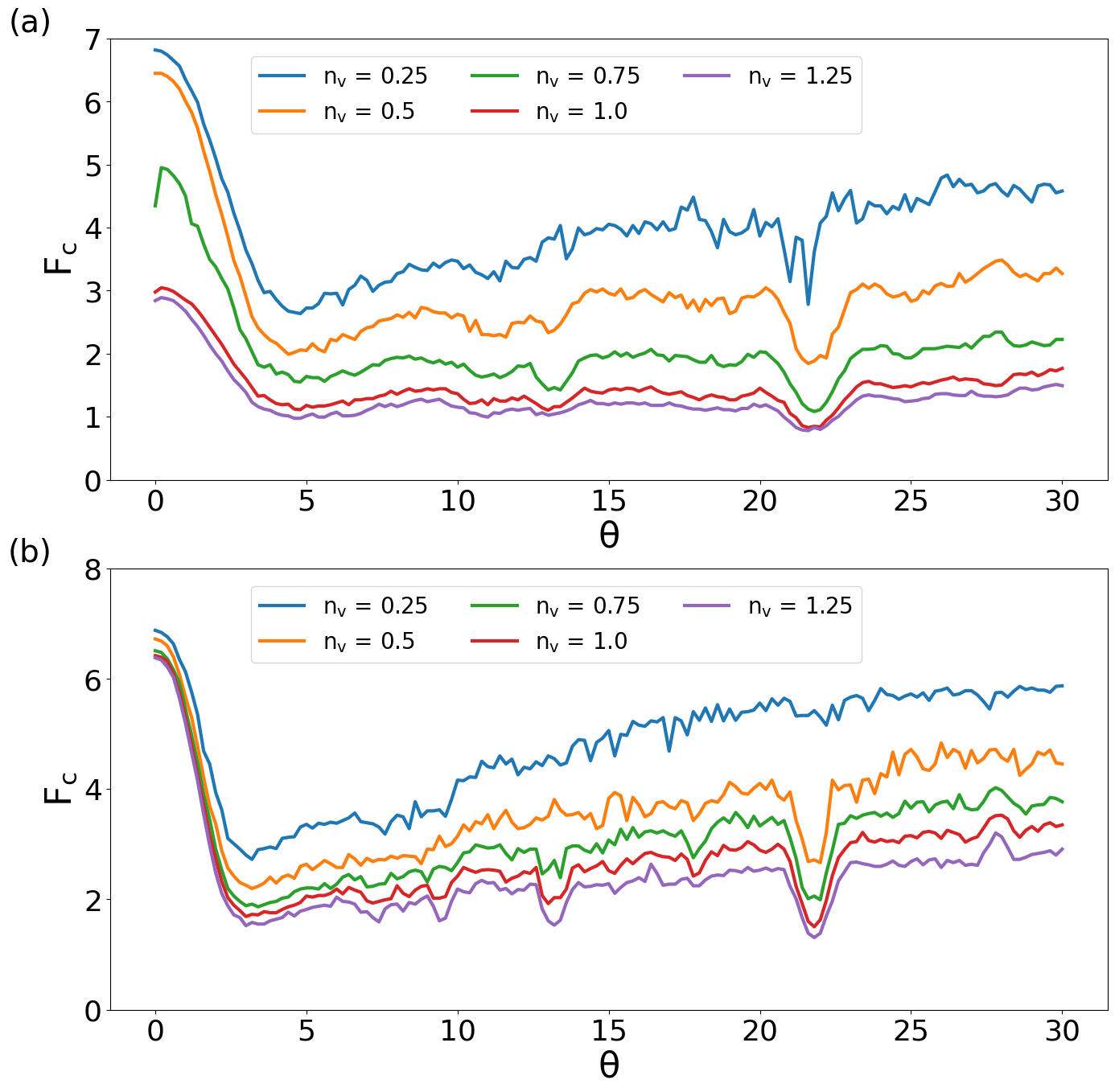}
\caption{
Critical current $F_c$ vs $\theta$ in samples of size (a) $L=16\lambda$ and (b) $L = 24\lambda$.}
\end{figure}
\begin{figure}
\includegraphics[width=0.88\columnwidth]{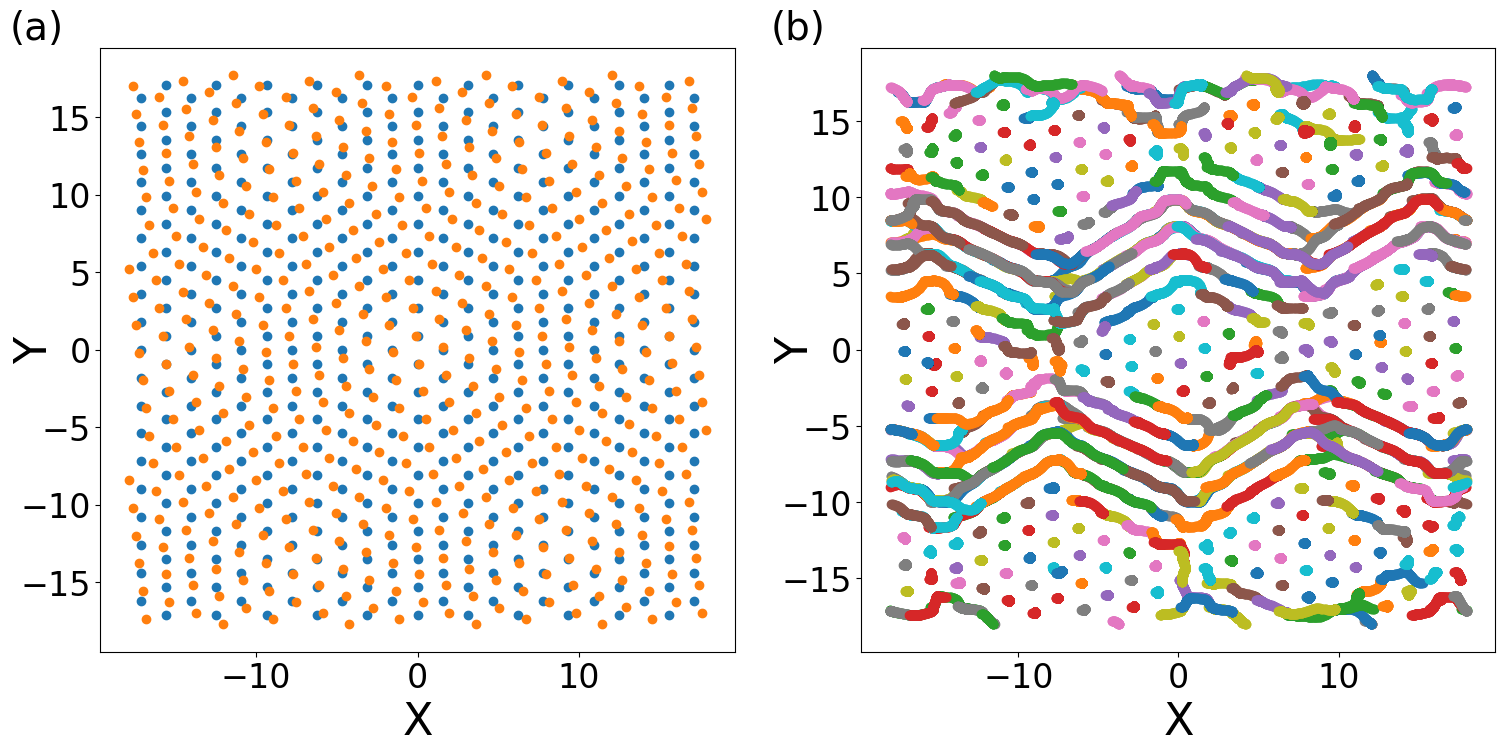}
\caption{
Vortex trajectories at $\theta = 6.6^\circ$, $n_v = 1.25$, and $F_d = 3.0$.}
\end{figure}
\begin{figure}
\includegraphics[width=0.88\columnwidth]{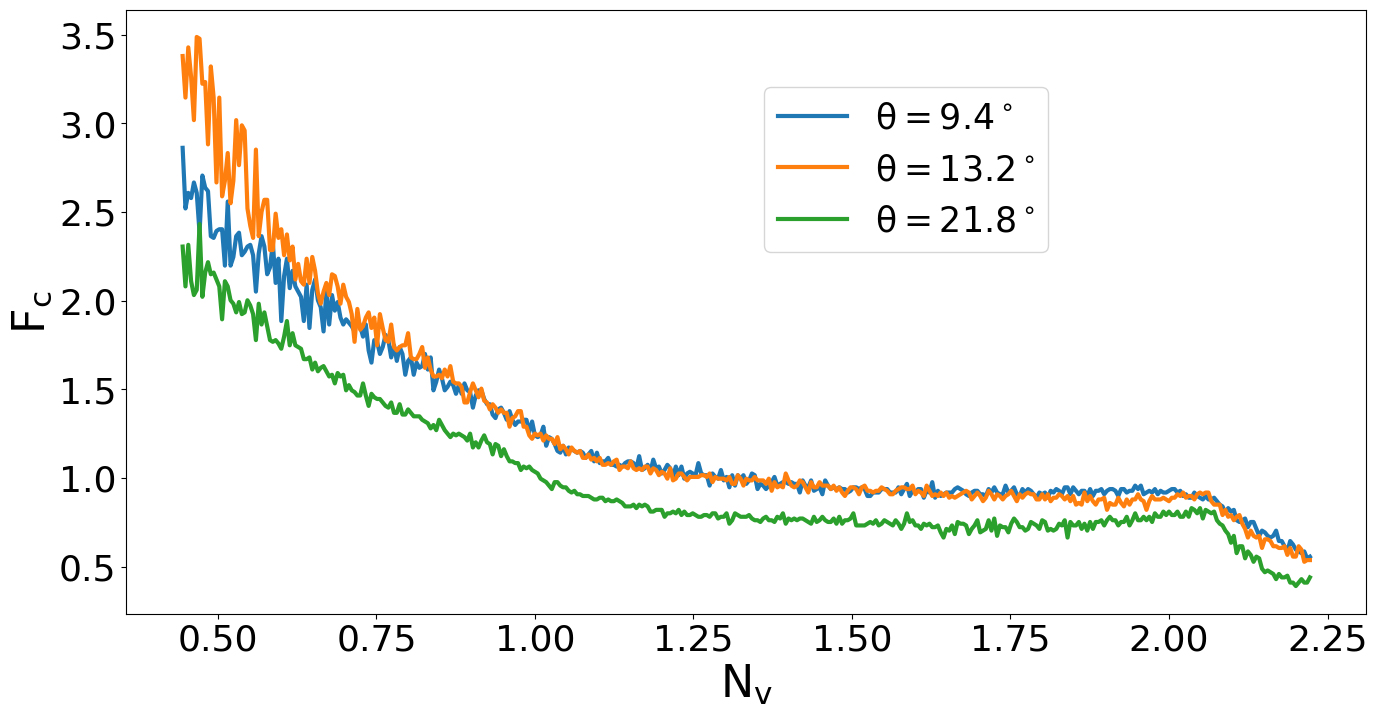}
\caption{
Critical current $F_c$ vs magnetic field $n_v$ for $\theta = 9.4^\circ$, 13.2$^\circ$, and 21.8$^\circ$. There is a drop in $F_c$ when $N_{v}/N_{p} > 2.0$. }
\end{figure}